\documentstyle[12pt,cite]{article}
\catcode`\@=11
\renewcommand{\thefootnote}{\fnsymbol{footnote}}

\begin{document}

\begin{titlepage}

{\hfill DFPD/93/TH/65}

\vspace{0.2cm}

{\hfill hept-th/9310149}

\vspace{1.2cm}

{\centerline{{\large \bf THE HIGGS MODEL FOR ANYONS AND LIOUVILLE
ACTION}\footnote[5]{Partly Supported by the European Community Research
Programme {\it Gauge Theories, applied supersymmetry and quantum
gravity}, contract SC1-CT92-0789}}}

\vspace{1.6cm}

\centerline{\large{\sc Marco}  {\sc  Matone}\footnote{e-mail:
matone@padova.infn.it, mvxpd5::matone}}

\vspace{0.1cm}

\centerline{\it Department of Physics ``G. Galilei'' -
Istituto Nazionale di Fisica Nucleare}
\centerline{\it University of Padova}
\centerline{\it Via Marzolo, 8 - 35131 Padova, Italy}

 \vspace{2.1cm}

\centerline{\large ABSTRACT}

\vspace{0.2cm}

We connect Liouville theory, anyons and Higgs model in a
purely geometrical way.

\end{titlepage}

\newpage

\setcounter{footnote}{0}

\renewcommand{\thefootnote}{\arabic{footnote}}

\renewcommand{\theequation}{\thesection.\arabic{equation}}
\newcommand{\mysection}[1]{\setcounter{equation}{0}\section{#1}}

\mysection{The Higgs model for anyons}

The main aim of this paper is to show that the Higgs model for
anyons and Liouville
theory are strictly related. The crucial point is based on the
following remark

\begin{itemize}

\item[{\bf a}.]{{\it The configuration space of $n$ anyons is the
 manifold of $n$ unordered points in} $\widehat{\bf C}={\bf C}\cup
\{\infty \}$
\begin{equation}
 M_n=(\widehat{\bf C}^n\backslash \Delta_n)/{Symm(n)},
\label{an1}\end{equation}
{\it with $\Delta_n$ the diagonal subset where two or more punctures
coincide.}}
\item[{\bf b}.]{{\it The
Liouville action on the Riemann sphere with $n$-punctures evaluated on the
classical solution is the K\"ahler potential for the natural metric
(the Weil-Petersson metric) on\footnote{The $PSL(2,{\bf C})$ group
reflects the M\"obius symmetry of the Riemann sphere. Thus one can
consider ${\cal M}_n$ as the configuration space for $n-3$ anyons
on the Riemann sphere with 3-punctures (for example at $0,1$ and
$\infty$).}}
\begin{equation}
{\cal M}_n=
(\widehat{\bf C}^n\backslash \Delta_n)/{Symm(n)}\times
PSL(2,{\bf C}).\label{aamodulisp}\end{equation}}
\end{itemize}

This remark implies that starting from anyons on $\widehat{\bf C}$ one can
recover the two-form associated to the
natural metric on the configuration space by first computing
the Poincar\'e metric $e^{\varphi_{cl}}$
on the punctured sphere and then, after evaluating
the Liouville action for $\varphi=\varphi_{cl}$, computing the
curvature two-form of the Hermitian line bundle
defined by classical action.

To understand the physical relevance of this remark
we recall that the quantum Hamiltonian for $n$ anyons is
proportional to the covariant Laplacian on $M_n$.

Actually, the connection between anyons and Liouville arises also
in considering the  critically coupled abelian Higgs model in
(2+1)-dimensions where the space $M_n$
plays a crucial role in the analysis of $n^{th}$ topological sector of the
theory.
Remarkably $f=2{\rm Re}\,\log \phi$, with
$\phi$ the Higgs field, satisfies the modified
(non covariant) Liouville equation
\begin{equation}
f_{z\bar z}=e^f-1,
\label{oid56}\end{equation}
which is the second Bogomol'nyi equation.
Notice that the non covariance (the $-1$
in (\ref{oid56})) is due to the Higgs mass.

Let us consider the Lagrangian of the Higgs model (see
\cite{samols,samols2,manton}
for details)
\begin{equation}
{\cal L}={1\over 2}D_\mu \phi {\overline {D^\mu \phi}}-
{1\over 4}F_{\mu\nu}F^{\mu\nu}-{1\over
8}\left(|\phi|^2-1\right)^2,
\qquad \phi=\phi_1+i\phi_2,
\label{higgs34}\end{equation}
where
$$
D_\mu\phi=(\partial_\mu-iA_\mu)\phi,\qquad  F_{\mu\nu}=
\partial_\mu A_\nu-\partial_\nu A_\mu,
$$
and the metric has signature $(1,-1,-1)$.

In the temporal gauge
the finiteness of the energy implies that at infinity
the Higgs field is a pure phase.
The magnetic flux through ${\bf R}^2$ is
\begin{equation}
\int F_{12}=2\pi n,\label{flux} \end{equation}
where $n$ is the winding number labelling the topological sectors
of the map
\begin{equation}
|\phi|: S^1_\infty\longrightarrow U(1).\label{odad}\end{equation}
 In the static
configuration $\dot A_i=0$, $\dot \phi=0$, the energy has the
lower bound $E\ge \pi |n|$. The critical case $E=\pi |n|$ arises
when the Bogomol'nyi equations
\begin{equation}
\left(D_1+{\rm sgn}(n) i D_2\right)\phi=0,
\qquad F_{12}+{\rm sgn}(n){1\over 2}\left(|\phi|^2-1\right)=0,
\label{bogomolnyi}\end{equation}
are satisfied.
It is crucial that in the $n^{th}$ topological sector the space of smooth
solutions is a
manifold $\widetilde M_n$ of complex dimension $n$.
In particular, each solution is uniquely specified by the $n$
unordered points $\{z_k\}$ where the Higgs field is zero
\cite{jt}. The same happens in considering the Liouville equation
for the Poincar\'e metric on Riemann surfaces. In particular,
due to the uniqueness of the solution of the Poincar\'e metric,
to each complex structure of the $n$-punctured Riemann sphere
(the unordered set of $n$ points) corresponds a solution of the
Liouville equation (see below for details).

Topologically
$\widetilde M_n$ coincides with $\widehat{\bf C}^n$.
It is a remarkable fact that the abelian Higgs model (almost) provides
a smooth metric and $U(1)$ gauge field on $\widetilde M_n$.
In particular the kinetic energy induces the metric
\begin{equation}
ds^2={1\over 2}\int_{\bf C}(d\phi_a d\phi_a +dA_idA_i).
\label{oix65}\end{equation}
The field evolution is described by geodesic motion on $\widetilde M_n$.

The $z_k$'s are
good coordinates only on the subspace $M_n$.
Good global coordinates on $M_n$ are provided by
the coefficients of the polynomial \cite{h}
\begin{equation}
P_n(z)=\sum_{k=0}^nw_kz^k\equiv \prod_{k=1}^n (z-z_k).\label{ply}
\end{equation}

Note that the field
evolution defines deformation of the complex structure of the punctured
sphere. Recently it has been shown that this deformation is strictly
related with Liouville theory (see below).

In \cite{samols} Samols introduced the following metric
on $M_n$
\begin{equation}
ds^2={1\over 2}\sum_{r,s=1}^n
\left(\delta_{rs}+2{\partial \bar b_r\over \partial z_s}\right)dz_rd\bar z_s,
\label{samols1}\end{equation}
where the $b_r$'s satisfy the equations
\begin{equation}
{\partial b_r\over \partial \bar z_s}={\partial \bar b_s\over
\partial z_r}.\label{accsrprlike}\end{equation}

\mysection{The Liouville equation}

Let us denote by $H$ the upper half plane and with $\Gamma$ a finitely
generated Fuchsian group. A Riemann surface isomorphic to the quotient
$H/\Gamma$ has the Poincar\'e metric $\hat g$ as the unique metric with scalar
curvature $R_{\hat g}=-1$ compatible with its complex structure. This implies
the uniqueness  of the solution of the Liouville equation on $\Sigma$. The
Poincar\'e metric on $H$ is
\begin{equation}
d{\hat s}^2={|dw|^2\over ({\rm Im}\,w)^{2}}.
\label{tyrop}\end{equation}
Note that $PSL(2,{\bf R})$ transformations are isometries of
$H$ endowed with the Poincar\'e metric.

An important property of $\Gamma$ is that it is isomorphic to the fundamental
group $\pi_1(\Sigma)$. Uniformizing groups admit the following structure.
Suppose $\Gamma$ uniformizes a surface of genus $h$ with $n$ punctures and $m$
elliptic points with indices $2\le q_1^{-1}\le q_2^{-1}\le \ldots\le
q_{m}^{-1}<\infty$. In this case the Fuchsian group is generated by $2h$
hyperbolic elements $H_1,\ldots,H_{2h}$, $m$ elliptic elements $E_1,\ldots,E_m$
and $n$ parabolic elements $P_1,\ldots,P_n$, satisfying the relations
\begin{equation}
E_i^{q_i^{-1}}=I,\qquad
\prod_{l=1}^mE_l\prod_{k=1}^nP_k\prod_{j=1}^h
\left(H_{2j-1}H_{2j}{H_{2j-1}^{-1}}{H_{2j}^{-1}}\right)=I,
\label{iodlkjnm}\end{equation}
where the infinite cyclicity of parabolic fixed point stabilizers is
understood.

Setting $w=J_H^{-1}(z)$ in (\ref{tyrop}),
where  $J_H^{-1}:\Sigma\to H$
is the inverse of the uniformization map, we get the Poincar\'e metric
on $\Sigma$
\begin{equation}
d{\hat s}^2=2{\hat g}_{z\bar z}|dz|^2=
e^{\varphi_{cl}(z,\bar z)}|dz|^2,\label{pncrsk}\end{equation}
where
\begin{equation}
e^{ \varphi_{cl}(z,\bar {z})}={|{J_H^{-1}(z)}'|^2\over({\rm Im}\,
J_H^{-1}(z))^2},\label{2prev}\end{equation}
which is invariant under $SL(2,{\bf R})$ fractional transformations
of $J_H^{-1}(z)$. Since
\begin{equation}
R_{\hat g}=-{\hat g}^{z\bar z}\partial_z\partial_{\bar z}
\log {\hat g}_{z\bar z},\qquad \hat g^{z\bar z}=2e^{-\varphi_{cl}},
\label{sc}\end{equation}
the condition $R_{\hat g}=-1$ is equivalent
to the Liouville equation
\begin{equation}
\partial_z\partial_{\bar {z}}\varphi_{cl}(z,\bar{z})=
{1\over 2}e^{ \varphi_{cl}(z,\bar{z})}.\label{1}\end{equation}

Eq.(\ref{2prev}) shows that from the explicit
expression of the inverse map we can find the dependence of
$e^\varphi_{cl}$ on the
moduli of $\Sigma$. Conversely we can express the inverse map (to within a
$SL(2,{\bf C})$ fractional transformation) in terms of $\varphi_{cl}$.
This follows from the Schwarzian equation
\begin{equation}\{J_H^{-1},z\}=
T^F(z),\qquad \label{4}\end{equation}
where
\begin{equation}
T^F(z)=\varphi_{cl}''-{1\over 2}(\varphi_{cl}')^2,
\label{stress1}\end{equation}
is the classical Liouville energy-momentum tensor and
\begin{equation}
\{f,z\}
={f'''\over f'}-{3\over 2}\left({f''\over f'}\right)^2=
-2(f')^{1\over 2}((f')^{-{1\over 2}})'',
\label{schrz}\end{equation}
is the Schwarzian derivative.

\mysection{The Riemann sphere}

 Here we now discuss basic geometrical results on the
Riemann sphere with $n$-punctures
\begin{equation}
\Sigma=\widehat
{\bf C}\backslash\{z_1,\ldots,z_n\},\qquad
\widehat {\bf C}\equiv {\bf C}\cup\{\infty\}.
\label{rsn}\end{equation}
Let $P_1,\ldots,P_n$, $n\ge 4$, be the set of parabolic generators
of $\Gamma$ satisfying the
constraint $P_1\cdot\cdot\cdot P_n=1$ and with the property that
their parabolic fixed points $\{w_1,\ldots,w_n\}\in
{\bf R}\cup\{\infty\}$ map onto $\{z_1,\ldots,z_n\}$.
The moduli space of $n$-punctured spheres is the
space of classes of isomorphic $\Sigma$'s, that is
\begin{equation}
{\cal M}_{n}=
\{(z_1,\ldots,z_{n})\in
\widehat{\bf C}^{n}|z_j\ne z_k\; {\rm for}\; j\ne k\}/Symm(n)\times
PSL(2,{\bf C}), \label{modulisp}\end{equation}
where ${Symm}(n)$ acts by permuting
$\{z_1,\ldots,z_n\}$ whereas $PSL(2,{\bf C})$ acts by linear fractional
transformations. By $PSL(2,\bf C)$ we can recover the `standard
normalization':  $z_{n-2}=0$, $z_{n-1}=1$ and $z_{n}=\infty $.
Furthermore, without loss of generality, we assume
that $w_{n-2}=0$, $w_{n-1}=1$ and
$w_n=\infty$.  For the classical Liouville tensor we have
\begin{equation}T^F(z)
=\sum_{k=1}^{n-1}\left({1\over 2(z-z_k)^2}+
{c_k\over z-z_k}\right),\qquad
\lim_{z\to \infty}T^F(z)={1\over 2z^2}+{c_n\over z^3}+
{\cal O}\left({1\over |z|^4}\right).\label{6}\end{equation}
Notice that $T^F$ is holomorphic
on $\Sigma=\widehat {\bf C}\backslash\{z_1,\ldots,z_n\}$.
Such a characteristic  of $T^F$ extends to higher
genus surfaces as well.
This follows from the Liouville equation.
Equivalently one can consider local univalence of the inverse map
which implies that ${\partial_zJ_H^{-1}}(z)\ne
0,\,\forall z\in \Sigma$, so that the Schwarzian derivative
of $J_H^{-1}$ is holomorphic on $\Sigma$.

Eq.(\ref{6}) implies the following constraints on the
{\it accessory parameters}
\begin{equation}
\sum_{k=1}^{n-1}c_k=0,\qquad \sum_{k=1}^{n-1}c_kz_k=1-n/2,\qquad
\sum_{k=1}^{n-1}z_k(1+c_kz_k)=c_n,\label{7}\end{equation}
so that  $c_1,\ldots,c_{n-3}$ can be considered as the basic set.
These parameters are functions on the space
\begin{equation} V^{(n)}=\{(z_1,\ldots,z_{n-3})\in
{\bf C}^{n-3}|z_j\ne 0,1; z_j\ne z_k,\; {\rm for}\; j\ne k\},
\label{8}\end{equation}
which is a covering of ${\cal M}_{n}$
whereas  the Teichm$\ddot{\rm u}$ller space $T_{n}$ is a covering of
$V^{(n)}$ whose transformations form a subgroup of the modular group.
Note that
\begin{equation}
{\cal M}_{n}\cong V^{(n)}/{Symm}(n),
\label{mdls}\end{equation}
where the action of $Symm(n)$ on $V^{(n)}$ is defined by comparing
(\ref{modulisp}) with (\ref{mdls}).
In particular, if a permutation
involves  at least one of the punctures between $0$, $1$ and $\infty$, then
we must perform a linear fractional transformation to recover the standard
normalization. This means that the last three punctures of the
transformed surface $\widetilde\Sigma$
must be $\tilde z_{n-2}=0$, $\tilde z_{n-1}=1$,
$\tilde z_{n}=\infty$. Thus if $\sigma_{k}\in Symm(n)$ interchanges
$z_k$ and $z_{k+1}$, the coordinate on $\widetilde\Sigma$ is
\begin{equation}
\tilde z= \left\{\begin{array}{cc} z, & k=1,\ldots,n-4,\\(z-z_k)/( 1-z_k),
& k=n-3, \\          1-z, & k=n-2, \\ z/(z-1), &  k=n-1.\end{array}  \right.
\label{newssurf}\end{equation}

\mysection{Liouville action and the Weil-Petersson metric}

Let us consider the Liouville
action on the Riemann spheres with $n$-punctures \cite{0}
\begin{equation}
S^{(n)}=\lim_{r\to 0}S^{(n)}_r=
\lim_{r\to 0}\left[\int_{\Sigma_r}
\left(\partial_z\varphi\partial_{\bar z}{\varphi}+e^{\varphi}\right)+
2\pi (n {\log} r+2(n-2){\log}|{\log}r|)\right],
\label{32}\end{equation}
$$
\Sigma_r=\Sigma\backslash\left(\bigcup_{i=1}^{n-1}
\{z||z-z_i|<r\}\cup\{z||z|>r^{-1}\}\right),
$$
where the field $\varphi$ is in the class of smooth functions on
$\Sigma$ with the boundary condition
given by the asymptotic behaviour of the classical solution
\begin{equation}
\varphi_{cl}(z) =\left\{\begin{array}{cc}
-2{\log}|z-z_k|-2 {\log}|{\log}|z-z_k||+{\cal O}(1),\;
& z\to z_k, \; k\ne n,\\
-2\log|z|-2 \log\log|z|+{\cal O}(1),\;
& z\to\infty, \end{array}\right.
 \label{aba}\end{equation}
 Eq.(\ref{32}) shows that already at the classical level
the Liouville action needs a regularization
whose effect is to cancel the contributions
coming from the non covariance\footnote{Recall that
$e^\varphi$ is a $(1,1)$-differential.} of $|\varphi_z|^2$.
This regularization provides a modular anomaly for the
Liouville action which  is strictly
related to the geometry of the moduli space \cite{asym}.
In particular, it turns out that
the Liouville action evaluated on the classical
solution $S^{(n)}_{cl}$ is not
invariant under the action of ${Symm}(n)$ \cite{asym}
$$
S^{(n)}_{cl}(z_1,\ldots,z_{n-3})-
S^{(n)}_{cl}(\sigma_{i,n}(z_1,\ldots,z_{n-3}))
$$
\begin{equation}
=\left\{\begin{array}{cc} 4\pi\sum_{k\ne i}
\log|z_k-z_i|-2\pi(n-4)\log|z_i(z_i-1)|,\; & i=1,\ldots,n-3, \\
4\pi\sum_{k=1}^{n-3} \log |z_k|,          \; & i=n-2, \\
   4\pi\sum_{k=1}^{n-3} \log |z_k-1|, \; &  i=n-1,\end{array}  \right.
\label{newssurf0}\end{equation}
where $\sigma_{i,n}\in {Symm}(n)$, $i\ne n$, is the transformation
interchanging the $i$ and $n$ punctures.
Furthermore, the asymptotic behaviour
of the classical Liouville action when the punctures coalesce is \cite{asym}
\begin{equation}
S^{(n)}_{cl}(z_1,\ldots,z_{n-3})=\left\{\begin{array}{cc} 2\pi
\log|z_k-z_i|+{\cal O}(1),\; & z_i\to z_k,\; k\ne n, \\
2\pi\log |z_i|+{\cal O}(1) \; & z_i\to\infty.\end{array}  \right.
\label{newssurf1}\end{equation}

It turns out that the Liouville action computed
on the classical solution is a continuously
 differentiable function on $V^{(n)}$ and \cite{0}
 \begin{equation}
-{1\over 2\pi}{\partial S^{(n)}_{cl}\over \partial z_k}=c_k,
\qquad k=1,\ldots,n-3,
\label{33}\end{equation}
where the $c_k$'s are the accessory parameters defined in (\ref{6}).

Notice that by (\ref{7}) and (\ref{33}) it follows that
$$
\sum_{k=1}^{n-3}{\partial S^{(n)}_{cl}\over \partial z_k}=
2\pi (c_{n-1}+c_{n-2}),\qquad
 \sum_{k=1}^{n-3}z_k{\partial S^{(n)}_{cl}\over \partial z_k}
=\pi(n-2+2 c_{n-1}),
$$
\begin{equation}
\sum_{k=1}^{n-3}z_k^2{\partial S^{(n)}_{cl}\over \partial z_k}
=2\pi \left(c_{n-1}-c_n+1+\sum_{k=1}^{n-3}z_k\right).
\label{gf09}\end{equation}
Since $S^{(n)}_{cl}$ is real, eq.(\ref{33}) yields
\begin{equation}
(\partial +{\overline\partial})S^{(n)}_{cl}=
 -2\pi\sum_{k=1}^{n-3}(c_kdz_k+{\overline c}_k d\bar z_k),
\label{nwis1}\end{equation}
and
\begin{equation}
{\partial c_j\over \partial z_k}={\partial c_k\over \partial z_j},
\qquad
j,k=1,\ldots,n-3,
\label{ohdqs98}\end{equation}
\begin{equation}
{\partial c_j\over \partial {\bar z_k}}=
{{\partial \bar c_k\over \partial {z_j}}},\qquad
j,k=1,\ldots,n-3.
\label{eqpara2}\end{equation}

We stress the strict similarity between
(\ref{eqpara2}) for the accessory parameters and the Samols equations
(\ref{accsrprlike}).

Another important result in \cite{0} is
 \begin{equation}
{\partial c_j\over\partial\bar z_k}={1\over 2\pi}\left \langle {\partial\over
\partial z_j}\,,{\partial\over \partial z_k} \right \rangle, \qquad
j,k=1,\ldots,n-3,\label{34}\end{equation}
where the brackets denote
the Weil-Petersson metric on the Teichm$\ddot{\rm u}$ller space
$T_{n}$ projected onto $V^{(n)}$.
Therefore by (\ref{33})
 \begin{equation}
{\partial^2 S^{(n)}_{cl} \over \partial z_j
\partial {\bar z_k}}=-
\left \langle {\partial\over
 \partial z_j}\,,{\partial\over \partial z_k
}\right \rangle ,\qquad j,k=1,\ldots,n-3,\label{35}\end{equation} that is
\begin{equation}\omega_{WP}= {i\over
2}{\overline\partial}{\partial}S^{(n)}_{cl}=-i\pi\sum_{j,k=1}^{n-3}
 {\partial c_k\over \partial {\bar z_j}}d\bar z_j\wedge d z_k,
\label{36}\end{equation} where $\omega_{WP}$ is the Weil-Petersson
two-form  on $V^{(n)}$.
Thus the Liouville action evaluated on the classical solution
is the K${\rm \ddot a}$hler  potential for the Weil-Petersson
two-form defining the symplectic structure on $V^{(n)}$.

\mysection{Geometric quantization}

The symplectic structure considered above allows us to consider
the following Poisson bracket relations \cite{t2}
\begin{equation}
\{c_j,c_k\}_{WP}=0,\qquad \{c_j,z_k\}_{WP}={i\over \pi}\delta_{jk},
\label{bb1aaa}\end{equation} where the brackets are defined with
respect to Weil-Petersson metric $\omega_{WP}$.
These relations suggest to performing
 the geometric quantization of the space ${\cal M}_n$.
To do this we must define a suitable
line bundle. Let us consider the function \cite{asym}
$$
f_{\sigma_{k,n}}={
\prod_{j\ne k,\,j=1}^{n-3}(z_j-z_k)^2\over (z_k(z_k-1))^{n-4}}, \qquad
k=1,\ldots,n-3,
$$
\begin{equation}
f_{\sigma_{n-2,n}}=\prod_{j=1}^{n-3}z_j^2,\qquad
f_{\sigma_{n-1,n}}=\prod_{j=1}^{n-3}(z_j-1)^2.
\label{ccycl1}\end{equation}
The extension by the composition
$f_{\sigma_1\sigma_2}=(f_{\sigma_1}\circ \sigma_2)f_{\sigma_2}$
defines a 1-cocycle $\{f_\sigma\}_{\sigma\in{Symm}(n)}$ of
${Symm}(n)$ \cite{asym}.
Let us now consider the holomorphic line bundle
\begin{equation}
{\cal L}_{n} =V^{(n)}\times {\bf C}/{Symm}(n),\label{lnbndl}\end{equation}
 on ${\cal M}_{n}$  where the action of $\sigma\in
{Symm}(n)$ is defined by
$(x,z)\to (\sigma x,f_\sigma (x)z)$, $x\in V^{(n)}$, $z\in {\bf C}$.
Since
\begin{equation}
\exp \left({ S^{(n)}_{cl}\circ \sigma
 \over \pi}\right)|f_\sigma|^2=\exp \left({ S^{(n)}_{cl}
\over\pi}\right),\label{37}\end{equation}
 it follows that
$\exp (S^{(n)}_{cl}/\pi)$ is a
  Hermitian metric in the line bundle ${\cal L}_{n}\to
{\cal M}_{n}$.
By (\ref{33})
 $\exp (S^{(n)}_{cl}/\pi)$ has connection form $-2\sum_i c_i d z_i$ and,
by (\ref{36}), curvature two-form
$-{ 2i\over  \pi}\omega_{WP}$ \cite{asym}.
The covariant derivatives are
\begin{equation}
{\cal D}_k =\partial_{z_k}-\partial_{z_k}{S^{(n)}_{cl}\over\pi}=
\partial_{z_k}+2c_k,\qquad
 \qquad {\overline{\cal D}}_k ={\partial}_{\bar z_k}.
\label{covq}\end{equation}

In the geometric quantization the Hilbert states
are sections of ${\cal L}_{n}$ annihilated by `half' of
the derivatives (polarization). The natural choice is
\begin{equation}
{\cal H}=\left\{ \psi\in {\cal L}_{n}| {\overline{\cal D}}_k
\psi=0\right\},
\label{hilbert1}\end{equation}
with inner product
\begin{equation}
\big<\psi_1|\psi_2\big>=
 {1\over n!(n-3)!}\int_{{\cal M}_n}
d(WP)
e^{-{S^{(n)}_{cl}\over \pi}}\overline \psi_1\psi_2,
\label{inner1}\end{equation}
where
\begin{equation}
d(WP)\equiv\left(\bigwedge^{n-3}{i\over 2}
\overline\partial\partial
S^{(n)}_{cl}\right),\label{wpv}\end{equation}
that by (\ref{36}) is the Weil-Petersson volume form.
An alternative to (\ref{wpv})
is to use the volume form $\bigwedge^{n-3} \omega$ where
\begin{equation}
\omega={i\over 2}\overline\partial\partial \log \det\,\left\|{\partial^2
S^{(n)}_{cl}\over \partial \bar z_j \partial z_k}\right\|.
\label{altern99}\end{equation}
However, in general, the correspondence principle can be proved only
if the scalar product is defined with respect to the
volume form associated with the
K${\rm \ddot a}$hler  potential \cite{berezin}. This means that in our
case we must use the Weil-Petersson volume form.

We conclude the discussion on the geometric quantization of ${\cal M}_n$
by noticing that there is an alternative
for the polarization choice in the geometric
quantization above, namely
\begin{equation}
{\cal H}=\left\{\widetilde \psi\in {\cal L}_{n}|
{\cal D}_k\widetilde \psi=0\right\},\label{polarization2}\end{equation}
that is, the  states and the classical Liouville action
are related by
\begin{equation}
\partial_{z_k}{S^{(n)}_{cl}}=\pi\partial_{z_k}\log
\widetilde\psi.\label{00}\end{equation}

\mysection{Anyons, Higgs and Liouville theory}

Let us make some remarks.
A crucial aspect that should be further investigated concerns the
classical-quantum interplay arising both in Liouville and
anyons theories\footnote{Notice that one must regard vortices
as unlabelled particles not
only quantum mechanically but also classically \cite{manton}.}.
In \cite{mma,mmb} it has been emphasized
that the regularization arising at the classical level
for the Liouville action is strictly related to the conformal properties
both of quantum and classical (Poincar\'e metric) Liouville operators.
A similar approach should be applied to anyons to provide a geometrical
interpretation of the statistics. As in the case of conformal
weights in 2D quantum gravity
\cite{mmb} we can consider anyons as elliptic points whose
ramification index fixes the statistics.

The approach considered here
 makes it possible \cite{mmc} to connect anyons theory with
the geometrical approach to quantum gravity recently considered
in \cite{mma,tak,mmb}.

Another interesting aspect is the connection between Liouville
and Higgs. This provides a way to consider the Higgs
model in a 2D gravity framework.

Finally we note that our results should be useful to investigate
the underlying
geometry\footnote{A crucial object in \cite{fm} is
the braid group
$B_n=\pi_1\left(M_n\right).$}
of the approach considered in \cite{fm}.

\vspace{1cm}

{\it Acknowledgements.} I would like to thank P.A. Marchetti for
stimulating discussions.

\end{document}